\documentclass[12pt]{article} 

\usepackage{amstex}
\usepackage{amssymb}

\usepackage{epsfig}

\usepackage{cite}

\begin{document}

\begin{titlepage}

\begin{flushright}
{\bf CERN-TH/96-176 \\
     UTHEP-96-0603}
\end{flushright}
 
\vspace{1mm}
\begin{center}
{\LARGE
Bhabha Process at LEP - Theoretical Calculations
}
\end{center}
 
\begin{center}
  {\bf S. Jadach}\\
   {\em Institute of Nuclear Physics,
        ul. Kawiory 26a, Krak\'ow, Poland}\\
   {\em CERN, Theory Division, CH-1211 Geneva 23, Switzerland,}\\
 {\bf M. Melles}\\
   {\em Department of Physics and Astronomy,\\
   The University of Tennessee, Knoxville, TN 37996-1200},\\
   {\bf B.F.L. Ward}\\
   {\em Department of Physics and Astronomy,\\
   The University of Tennessee, Knoxville, TN 37996-1200\\
   and 
   SLAC, Stanford University, Stanford, CA 94309}\\
   {\bf S.A. Yost }\\
   {\em Department of Physics and Astronomy,\\
   The University of Tennessee, Knoxville, TN 37996-1200}
\end{center}

\vspace{1mm}
\begin{abstract}
In this contribution we give a short overview of the situation in the
precision calculation of the Bhabha process and we present a preliminary
numerical result on the second-order sub-leading correction to
the small angle Bhabha process.
\end{abstract}
 
\begin{center}
{\it Presented by S. Jadach at Zeuthen Workshop, 
     April 1996, Rheinsberg, Germany}
\end{center}

\vspace{10mm}
\renewcommand{\baselinestretch}{0.1}
\footnoterule
\noindent
{\footnotesize
\begin{itemize}
\item[${\S}$]
Work supported in part by the US DoE contract DE-FG05-91ER40627,
Polish Government grants KBN 2P30225206, 2P03B17210,
the European Union under contract No. ERB-CIPD-CT94-0016
and Polish-French Collaboration within IN2P3.
\end{itemize}
}

\begin{flushleft}
{\bf CERN-TH/96-176\\
     UTHEP-96-0601\\
     June  1996 }
\end{flushleft}
 
\end{titlepage}

\def\Order#1{${\cal O}(#1$)}
\def\half{ {1\over 2} }
\def\alf1{ {\alpha\over\pi} }
\def\bbe{\bar{\beta}}

\section{Introduction}

The Bhabha scattering process consists in fact of two distinct processes
(especially at the Z peak): one is the Small Angle Bhabha (SABH) process
below about $6^\circ$ in the scattering angle,
which is dominated by the gamma $t$-channel exchange 
and another one, the Large Angle Bhabha (LABH) process
above $6^\circ$, which gets important contributions from
various $s$-channel (annihilation) exchanges.
The SABH process is employed almost exclusively to determine the luminosity
in the $e^+e^-$ colliders, using dedicated luminometer sub-detectors.
The LABH provides input data for precision electroweak tests of the
Standard Model (SM), in particular the electron partial width $\Gamma_e$
of the Z boson.
In this presentation we shall concentrate mainly on the SABH process.
At LEP at $\sqrt{s}=M_{\rm Z}$, in the $1^{\circ}$--$3^{\circ}$
angular range it gives about four times more events than Z decays.
It is therefore ideally suited for precise measurements of the luminosity
from the point of view of statistical error.
Even more important, it is dominated by ``known physics'', that is by
$t$-channel exchange of a photon -- it is therefore calculable
from ``first principles'', i.e. from the Lagrangian of
the Quantum Electrodynamics (QED) with the standard Quantum Field Theory
methods, Feynman diagrams, etc.

\subsection{Theoretical error in the luminosity measurement}

At present, the luminosity measurement at LEP using the SABH process has
a very small statistical and experimental systematic error, typically
$0.07-0.15\%$.
The uncertainty of the theoretical calculation of the SABH process
has to be combined with this error.
It is called the ``theoretical error'' (the theory uncertainty)
of the luminosity.
It was last year reduced to $0.16\%$ \cite{th-95-38} 
and is now at the level of $0.11\%$ 
\cite{YR-96-01-bhabhaWG,PL-bhabhaWG:1996}:
in spite of the progress,
it is still a dominant component of the total luminosity error.
This error enters into that of the total cross section
measured at LEP.
The experimental precision of
the so-called invisible width (number of neutrinos) is strongly
dependent on the precision of the luminosity measurement.
The other quantities used for tests of the SM are also affected.
In Table~\ref{STMOD} we show the influence of the luminosity
error on the LEP measurable used in the test of the SM.

\begin{table*}[!ht]
\centering
\begin{tabular}                            {cccc}
\hline
          &
\multicolumn{3}{c}{Theoretical luminosity error} \\ \hline
   & 0.16\% &  0.11\% & 0.06\% 
\\
\hline
$m_Z$~[GeV]
   & $ 91.1884\pm0.0022$ & $ 91.1884\pm0.0022$ & $91.1884\pm0.0022$ \\
$\Gamma_Z$~[GeV] 
   & $ 2.4962\pm0.0032$  & $ 2.4962\pm0.0032$  & $2.4961\pm0.0032$  \\
$\sigma^0_h$~[nb] 
   & $ 41.487\pm0.075$   & $ 41.487\pm0.057$   & $41.487\pm0.044$   \\
$R_l$
   & $ 20.788\pm0.032$   & $ 20.787\pm0.032$   & $20.786\pm0.032$   \\
$A_{FB}^{0,l}$ 
   & $ 0.0173\pm0.0012$  & $ 0.0173\pm0.0012$  & $0.0173\pm0012$    \\
\hline
$\Gamma_{had}$~[GeV]
   & $ 1.7447\pm0.0030$  & $ 1.7447\pm0.0028$  & $1.7446\pm0.0027$ \\
$\Gamma_{ll}$~[MeV]
   & $ 83.93\pm0.13$     & $ 83.93\pm0.13$     & $83.93\pm0.12$    \\
$\sigma^0_{ll}$~[nb]
   & $ 1.9957\pm0.0044$  & $ 1.9958\pm0.0038$  &$ 1.9959\pm0.0034$ \\
${\Gamma_{had}}/{\Gamma_Z}$~[\%]
   & $69.90\pm0.089$     & $ 69.90\pm0.079$    &$ 69.89\pm0.072$   \\
${\Gamma_{ll}}/{\Gamma_Z}$~[\%]
   & $3.362\pm0.0037$    & $ 3.362\pm0.0032$   &$ 3.362\pm0.0028$  \\
$\Gamma_{inv}$~[MeV]
  & $ 499.9\pm2.4$       & $ 499.9\pm2.1$      & $499.9\pm1.9$     \\
${\Gamma_{inv}}/{\Gamma_{ll}}$~[\%]
  & $5.956\pm0.030$      & $ 5.956\pm0.024$    &$ 5.956\pm0.020$   \\
$N_{\nu}$
  & $ 2.990\pm0.015$     & $ 2.990\pm0.013$    & $2.990\pm0.011$   \\
\hline
\end{tabular}
\caption{\small\sf
Line shape and asymmetry parameters from 5-parameter fits to the data of the
four LEP1 experiments, made with a theoretical luminosity error of 0.16\%, 
0.11\% and 0.06\%~\protect\cite{teubert-privcom:1995}.
In the lower part of the table also derived parameters are listed.
}
\label{STMOD}
\end{table*}

Obviously it would be worthwhile to lower the theoretical uncertainty
in the calculation of the SABH cross section below the future,
ultimate, experimental precision of LEP experiments, which will probably
reach 0.05\%.
From the beginning of the LEP operation both experimental and theoretical
components in the error of the luminosity went from the level of 2\% to 0.1\%.
Why was it always difficult to reduce the theoretical error even further?
The main obstacles were the need for  non-trivial calculations of the
higher-order contributions and the complicated Event Selection (ES)
in the actual measurement.
Due to the complicated ES, the phase-space boundaries in the calculation
of the SABH cross section are too complicated for any analytical calculation.
The calculation has to be numerical, the best being
in the form of the Monte Carlo event generator MCEG.
The theoretical calculation would be completely useless if
in the calculation
of the SABH cross section we did not control its ``technical precision'',
corresponding to {\em all} possible numerical uncertainties.
The control over the technical precision is probably the most difficult
and labour-consuming part of the enterprise.

\subsection{BHLUMI MC event generator}

In the recent years the LEP collaborations have
used the BHLUMI Monte Carlo event generator to calculate the SABH cross section
for any type of experimental ES.
The program, originally written in 1988 \cite{bhlumi1:1989},
was published in 1992 \cite{bhlumi2:1992} with
the first-order QED matrix element ${\cal O}(\alpha^1)_{exp}$
(exponentiation according to the Yennie-Frautschi-Suura theory)
and its matrix element was recently upgraded 
by means of adding the missing second-order in the Leading-Logarithmic
approximation \cite{bhlumi4:1996}.
BHLUMI provides multiple soft and hard photons in the complete
phase-space in all versions.
The multi-photon integration over multi-photon phase-space 
remains essentially unchanged in BHLUMI since the first version.
Gradual improvements concern mainly the matrix element.
More and more cross-checks are built up in order to better determine
its technical precision, see \cite{th-95-38,th-96-156}.

\subsection{Importance of the various QED corrections}

\begin{table*}[!ht]
\centering
\newcommand{\lstrut}{{$\strut\atop\strut$}}
\newcommand{\mystrut}{{\hbox{\rule[-3mm]{0mm}{8mm}}}}
\newcommand{\perm}{{\times 10^{-3}}}
\begin{tabular}{lrcccc}
\hline
    \multicolumn{2}{c}{ }
  & \multicolumn{2}{c}{  $\theta_{min}=30$~mrad  }
  & \multicolumn{2}{c}{ $\theta_{min}=60$~mrad   }
\\  \hline
    \multicolumn{2}{c}{ }
  & LEP1 &  LEP2 
  & LEP1 &  LEP2 
\\  \hline \mystrut 
${\cal O}(\alpha L    )$     &    $  \alf1 4L$         
       &  $137\perm$   &  $152\perm$   &  $150\perm$   &  $165\perm$  
\\  \hline \mystrut 
${\cal O}(\alpha      )$     &    $2 \half \alf1 $    
       &  $2.3\perm$  &  $2.3\perm$    &  $2.3\perm$   &  $2.3\perm$  
\\ \hline \mystrut 
${\cal O}(\alpha^2L^2 )$     &    $  \half \left(\alf1 4L\right)^2 $ 
       &  $9.4\perm$   & $11\perm$   & $11\perm$       & $14\perm$   
\\ \hline \mystrut 
${\cal O}(\alpha^2L )$       &    $ \alf1 \left(\alf1 4L\right) $ 
       &  $0.31\perm$  & $0.35\perm$  & $0.35\perm$    & $0.38\perm$  
\\ \hline \mystrut 
${\cal O}(\alpha^3L^3)$      &    $ {1\over 3!}\left(\alf1 4L\right)^3 $ 
       &  $0.42\perm$  & $0.58\perm$  & $0.57\perm$    & $0.74\perm$  
\\ \hline
\end{tabular}
\caption{\small\sf
\baselineskip=12pt
The canonical coefficients indicating the generic magnitude of various
leading and sub-leading contributions up to third-order.
The big-log $L=\ln(|t|/m_e^2)-1$ 
is calculated for $\theta_{min}=30$~mrad and $\theta_{min}=60$~mrad
and for two values of the centre of mass energy:
at LEP1 ($\protect\sqrt{s}=M_Z$), where
the corresponding values of
$|t|=(s/4)\theta_{min}^2$ are 1.86 and  7.53~GeV$^2$, and
at a LEP2 energy ($\protect\sqrt{s}=200$~GeV), 
where the corresponding value of $|t|$ are 9 and 36~GeV$^2$, respectively. 
}
\label{tab:canonical-coeff}
\end{table*}

The electron mass is very small and the Leading-Logarithmic approximation
in terms of the big logarithm $L=\ln(|t|/m_e^2)-1$ is a very useful tool.
In Table~\ref{tab:canonical-coeff} we show numerical values of the
``canonical coefficients'' for various LL and sub-leading QED radiative
corrections. 
As we see from the table, for a precision of order 0.5\% it is 
enough to include the entire first-order \Order{\alpha} 
and   the second-order leading-log \Order{\alpha^2L^2},
while at the present precision, of order 0.05\%--0.10\%, it is necessary
to have control over \Order{\alpha^2L} and \Order{\alpha^3L^3}.

\subsection{Outline}

In the following section we shall briefly summarize
the results of the Bhabha Working group
in the recent LEP2 workshop, which led to a new lower precision estimate of
0.11\% of the theoretical uncertainty in the luminosity.
In the next section we shall give a glimpse of the recent
calculation of the critical second-order sub-leading correction
to the SABH process, which opens the route towards a theoretical precision
below 0.10\%.

\section{Bhabha Working group of LEP2 workshop}

\begin{figure*}[!ht]
\centering
\setlength{\unitlength}{0.1mm}
\begin{picture}(1600,1600)
\put( -20, 800){\makebox(0,0)[lb]{
\epsfig{file=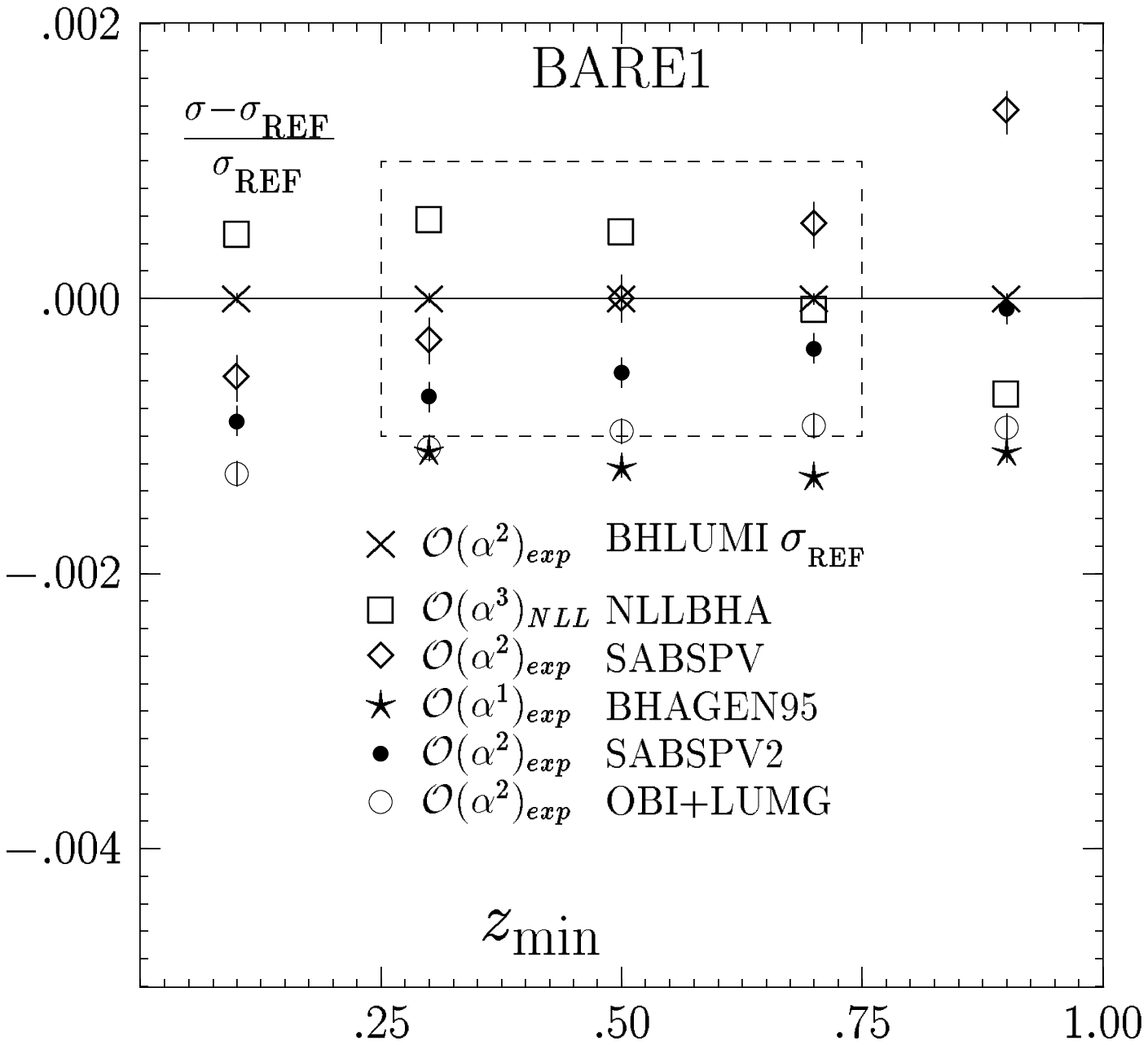,width=82mm,height=80mm}
}}
\put( 770, 800){\makebox(0,0)[lb]{
\epsfig{file=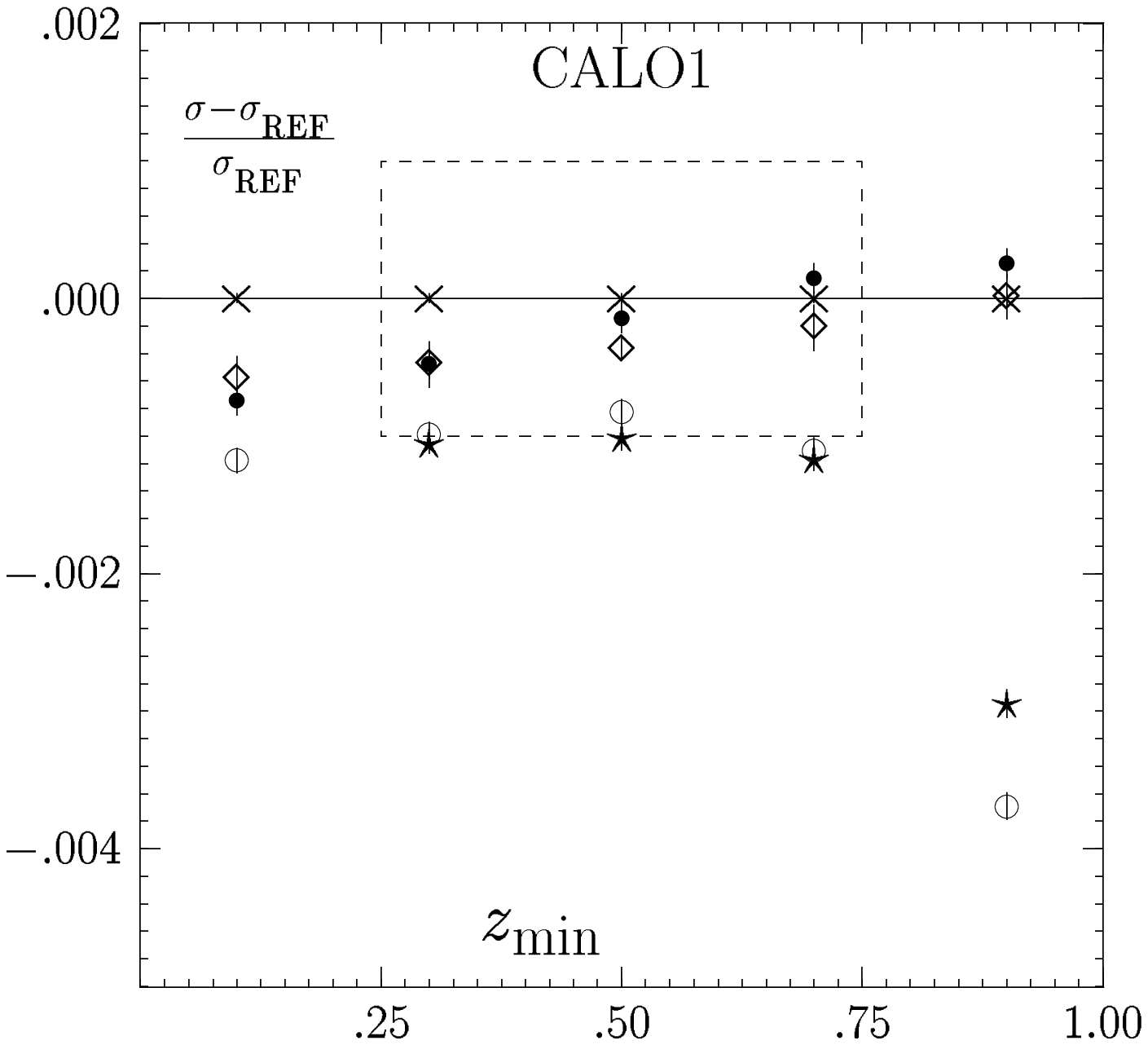,width=82mm,height=80mm}
}}
\put(-20,  00){\makebox(0,0)[lb]{
\epsfig{file=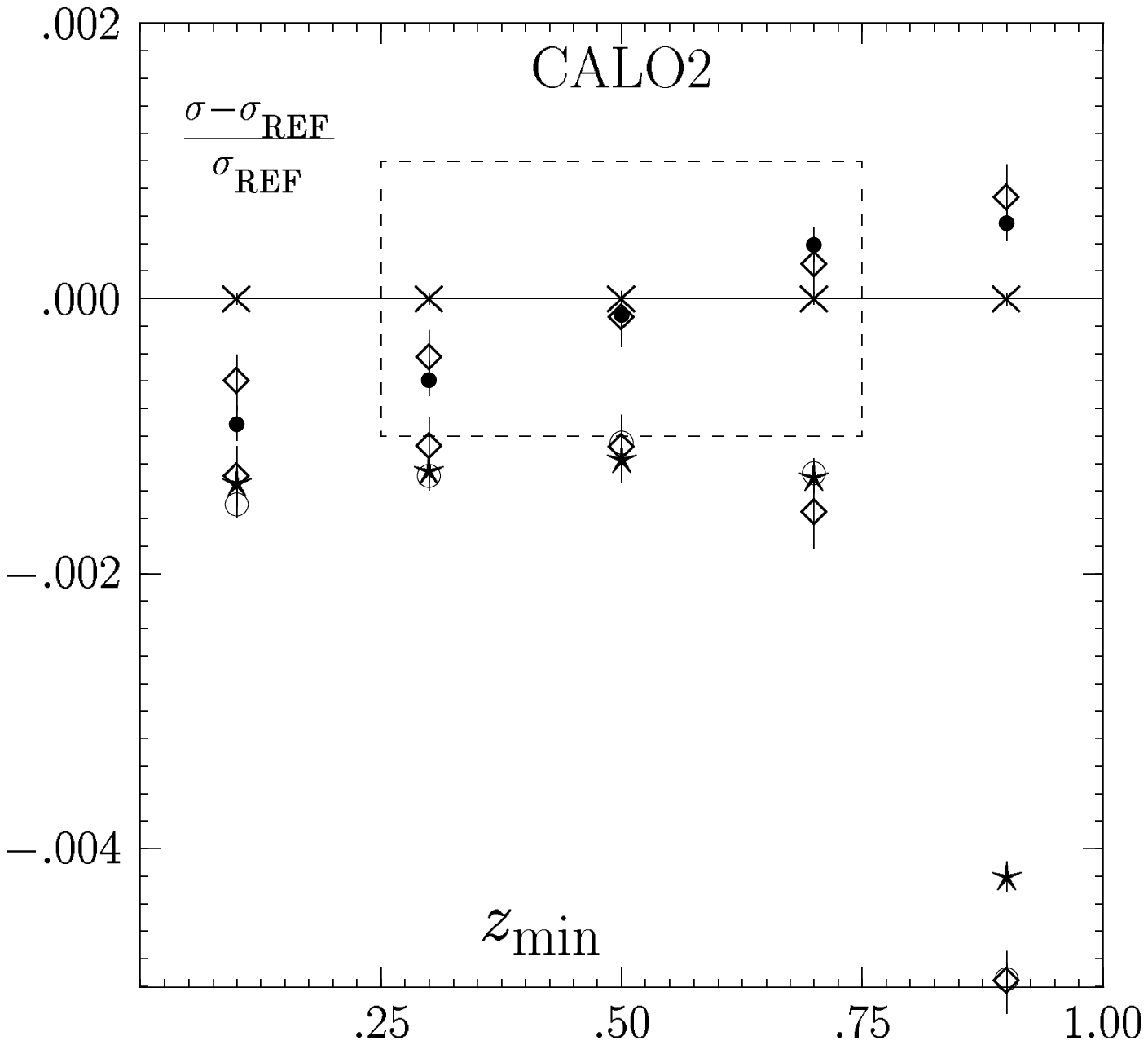,width=82mm,height=80mm}
}}
\put( 770,  00){\makebox(0,0)[lb]{
\epsfig{file=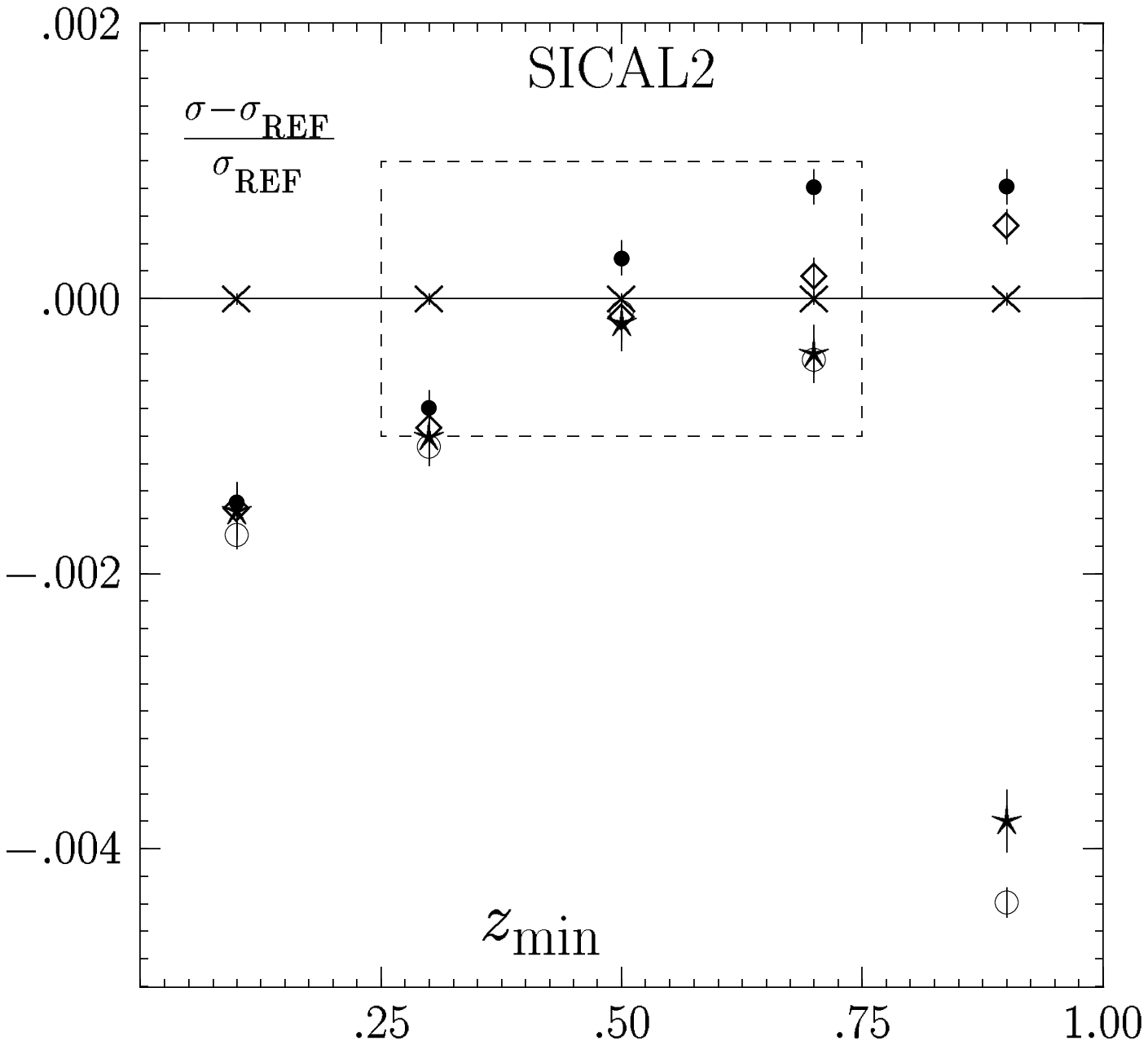,width=82mm,height=80mm}
}}
\end{picture}
\vspace{-9mm}
\caption{\small\sf
 Monte Carlo results for the symmetric Wide-Wide ES's 
 BARE1, CALO1, CALO2 and SICAL2, 
 for matrix elements beyond first-order.
 Z exchange, up-down interference and vacuum polarisation
 are switched off.
 The centre of mass energy is $\protect\sqrt{s}=92.3$~GeV.
 In the plot, the  ${\cal O}(\alpha^2)^{YFS}_{exp}$ cross section
 $\sigma_{_{\rm{BHL}}}$ from BHLUMI 4.02.a
 is used as a reference cross section.
}
\label{fig:sical92-alf2exp}
\end{figure*}
The main aim of the 1995 Bhabha Working group was to compare different
QED calculations for the SABH and LABH processes, in order to verify
and/or improve precision estimates for these calculations,
for both LEP1 and LEP2 applications.
It has to be stressed that it was really the first systematic and organized
example of such a comparison,
although the first step in this direction was already taken 
in Ref.~\cite{YR-95-03-partIII}.
The QED calculations for the SABH process, apart from BHLUMI, were
provided by four groups of authors: 
the names of the SABH programs/calculations 
are\footnote{Here we gave only one reference
      per program, see \cite{YR-96-01-bhabhaWG} for an exhaustive 
      list of relevant references.}
SABSPV \cite{sabspv:1995}, BHAGEN95 \cite{caffo:1994} 
and NLLBHA \cite{arbuzov:1995a}.
Among these four, BHLUMI represents a full-scale event generator,
SABSPV and BHAGEN95 are MC programs providing the total cross section
for arbitrary ES, and NLLBHA is a semi-analytical program able to calculate
the total cross section only for certain special (unrealistic) ES's.
The QED matrix element in all the calculations includes complete
first-order and second-order in the leading-logarithmic approximation,
with the notable exception of NLLBHA, which features in addition,
the second-order next-to-leading-log corrections.
All calculations feature some kind of exponentiation and part or all
of the third-order LL corrections.

In order to be able to better understand the differences between the four
calculations the WG participants agreed on four examples of
the ``standard event selection'' (SES) named BARE1, CALO1, CALO2, SICAL2.
These vary from the simple and unrealistic BARE1 (for which NLLBHA can
provide results) to the more sophisticated CALO1 and CALO2, ending with
SICAL2, which is very close to a typical experimental ES at LEP.
All four SES's, represent the ``double tag'' type of event selection,
that is both $e^+$ and $e^-$ have to be observed in forward/backward direction
with a certain minimum energy and minimum scattering angle.
For BARE1 the observed scattered objects are just ``bare'' $e^+$ and $e^-$
while accompanying bremsstrahlung photons are ignored completely.
This is unrealistic because all LEP luminometers are of the calorimetric
type, i.e. they combine
in the final state the $e^\pm$ with the photons that are close to them
into single objects, ``clusters''. 
The scattering angle of the cluster (which is the angle of its centre)
is required to be above a certain minimum angle,
below a certain maximum angle,
and the total energy of the cluster has to be above a certain threshold,
typically half of beam energy.
Note that the most sophisticated silicon luminometers
are not able to distinguish electrons and photons at all!
All three SES's, CALO1, CALO2 and SICAL2, are calorimetric and differ
in the way the cluster is defined.
CALO1 uses an angular cone around the directions of $e^\pm$ in order to
associate photons with the $e^\pm$ while CALO2 defines the cluster in terms
of a plaquette in the $(\theta,\phi)$ plain centred around $e^\pm$.
CALO2 and CALO1 still make a distinction between electrons and photons.
SICAL2 is completely charge-blind and forms a cluster around the most
energetic $e^\pm$ or photon exactly as in the silicon luminometer
of ALEPH or OPAL.
Finally, let us note that all four ES's are in two versions,
``symmetric'' and ``asymmetric''.
In the symmetric version the minimum and maximum value of the scattering angle
$\theta$ in the forward and backward hemisphere are the same;
in the asymmetric version, they are not.
The real experimental ES is asymmetric in order to eliminate
effects due to the geometrical uncertainty of the interaction point.
The interested reader will find in Ref.~\cite{YR-96-01-bhabhaWG} a 
more detailed description of the above four ES's.

The comparisons between the four calculations started with the warming-up
exercise in which all four groups have calculated {\em the same} first-order 
${\cal O}(\alpha^1)$ cross for all four SES's (except NLLBHA, which
is able to provide a cross section only for BARE1).
After some adjustment of the matrix element 
(Z-exchange and vacuum polarization were switched off) and debugging
of the programs for SES's very good agreement was obtained.
The cross sections at ${\cal O}(\alpha^1)$ agreed to within 0.03\%.
The calibration test was passed successfully for all SES's
including the realistic SICAL2.
In the above test the matrix element was exactly the same
and what was really seen as a difference was the pure technical
precision.
It would be ideal to extend this kind of test to second-order,
but here we could not do the same.
The matrix elements are not compatible, because in most of the calculations
the second-order sub-leading terms are incomplete.
The procedure of the phase-space integration over the two real photons
is for SABSPV and BHAGEN95 inherently tied up with adding second-order
LL correction and/or exponentiation.

In the next step the comparison of all calculations was attempted
for all four SES's for the matrix element ``beyond-first-order'',
using the best available QED matrix element for a given program.
In order to minimize the possible differences, Z-exchange and vacuum
polarization were temporarily switched off.
The results of the calculations are shown in Fig.~\ref{fig:sical92-alf2exp}.
All cross sections are compared with the BHLUMI cross section, which is
used as a reference\footnote{A table of the absolute cross
   sections is given in Ref.~\cite{YR-96-01-bhabhaWG}.}.
The differences with BHLUMI are plotted as a function of the dimensionless
energy-cut parameter $z_{\min}$. 
The minimum energy of $e^\pm$ in units of beam energy is approximately
$z_{\min}$, and the actual definition of $z_{\min}$ is slightly
different for each ES, see Ref.~\cite{YR-96-01-bhabhaWG} for more details.

As we see in Fig.~\ref{fig:sical92-alf2exp}, we have included 
in the comparisons another cross section calculated by means of BHLUMI,
which is referred to as being computed by OLDBIS $+$ LUMLOG.
The corresponding method of adding the second-order LL correction to
the first-order cross section (for arbitrary ES) was described in 
Ref.~\cite{th5995:1991} and the tools to calculate it are included
in the BHLUMI as separate sub-programs.
These tools are: (a) the first-order event generator OLDBIS 
and (b) the LL MC event generator LUMLOG, which generates photons 
in the strictly collinear approximation.
The comparison of BHLUMI with OLDBIS $+$ LUMLOG was
used in Refs.~\cite{th6118:1991,th-95-38} in order to estimate
the technical precision of BHLUMI and the missing higher-order and/or
sub-leading contributions.
In fact the OLDBIS $+$ LUMLOG recipe is quite similar
to the SABSPV calculation.
The main difference is that while the OLDBIS $+$ LUMLOG prescription
for combining \Order{\alpha} cross section with 
the \Order{\alpha^2}$_{{\rm LL}}$ was ``additive''
\begin{equation}
  {\cal O}(\alpha^1)
+ \{ {\cal O}(\alpha^2)_{{\rm LL}} - {\cal O}(\alpha^1)_{{\rm LL}} \}
\end{equation}
the SABSPV recipe is ``multiplicative''
\begin{equation}
  {\cal O}(\alpha^2)_{{\rm LL}} \times
  \Bigg\{ 1 +{{\cal O}(\alpha^1) - {\cal O}(\alpha^1)_{{\rm LL}} 
                                                      \over {\rm Born}}
  \Bigg\}.
\end{equation}
In both cases the ${\cal O}(\alpha^3)_{{\rm LL}}$ can be easily
used instead of ${\cal O}(\alpha^2)_{{\rm LL}}$.
Finaly, in Fig.~\ref{fig:sical92-alf2exp} we also show
cross section denoted SABSPV2 (dots) 
which is obtained according to multiplicative
prescription of SABSPV but using the cross sections from OLDBIS 
and LUMLOG\footnote{The slight difference between SABSPV and SABSPV2
  is of pure technical origin. 
  The  SABSPV2 result is not included in 
  Refs. \protect\cite{YR-96-01-bhabhaWG,PL-bhabhaWG:1996} --
  was obtained after the LEP2 workshop.}.

As we see in Fig.~\ref{fig:sical92-alf2exp}, there are distinct regularities
in the results.
The OLDBIS $+$ LUMLOG additive ansatz coincides extremely well with
the results of BHAGEN95, because the latter is also based on the additive
recipe.
The difference between BHLUMI and the result of additive recipes is, 
consistently with the older papers \cite{th6118:1991,th-95-38},
within 0.15\% for the experimentally interesting range 
$0.25 < z_{\min} < 0.75$.
The result of the multiplicative prescription of SABSPV agrees
with BHLUMI better; in fact, it stays for the same $z_{\min}$ range
within 0.1\% ``permille box''.
This is an interesting result if we remember that 
the multiplicative prescription
is a little better from the point of view of physics,
since it can effectively account for the part of the phase 
space with one real photon collinear
and one real photon acollinear to $e^\pm$.
The additive prescription simply ignores such configurations.
Another encouraging result, albeit only for the unrealistic BARE1 ES,
is the good agreement of the NLLBHA result with BHLUMI.
If taken seriously it would mean that the missing \Order{\alpha^2}
next-to-leading-log (NLL) contribution in BHLUMI is indeed below 0.1\%.
One would really need a similar result for more realistic ES's.
The above comparisons were also extended to the asymmetric version of the SES's
and they were also repeated for the situation when the  Z-exchange
and vacuum polarization were restored.
As can be seen in Ref.~\cite{YR-96-01-bhabhaWG},
similar agreements were obtained.
For LEP2 the agreements are slightly worse but definitely better than
0.20\%.

\begin{table*}[!ht]
\centering
\begin{tabular}{llll}
\hline
 & \multicolumn{2}{c}{LEP1} & LEP2 \\
\hline
  Type of correction/error    
& Ref.~\protect\cite{th-95-38}
& Present 
& Present     \\
\hline
(a) Missing photonic ${\cal O}(\alpha^2 L)$ &
    0.15\%      & 0.10\%    & 0.20\% \\
(a) Missing photonic ${\cal O}(\alpha^3 L^3)$ &
    0.008\%     & 0.015\%    & 0.03\%\\
(c) Vacuum polarization &
    0.05\%      & 0.04\%    & 0.10\% \\
(d) Light pairs &
    0.01\%      & 0.03\%    & 0.05\%\\
(e) Z-exchange  &
    0.03\%      & 0.015\%   &  0.0\%\\
\hline
    Total  &
    0.16\%      & 0.11\%    & 0.25\%\\
\hline
\end{tabular}
\caption{\small\sf
Summary of the total (physical+technical) theoretical uncertainty 
for a typical
calorimetric detector. 
For LEP1, the above estimate is valid for the angular range
within   $1^{\circ}$--$3^{\circ}$, and 
for  LEP2  it covers energies up to 176~GeV, and 
angular range within $1^{\circ}$--$3^{\circ}$ and $3^{\circ}$--$6^{\circ}$ 
(see the text for further comments). 
}
\label{tab:total-error-lep1}
\end{table*}

\subsection{New theoretical error estimate}

Following the above results the new estimate of the total theoretical
error for the BHLUMI cross section was obtained.
It is summarized in Table~\ref{tab:total-error-lep1} with the various
components of the theoretical error listed.
The older results of Ref.~\cite{th-95-38} and
the conservative projection of the theoretical error for LEP2 are also given.
The main progress is done for the missing photonic ${\cal O}(\alpha^2 L)$,
which was reduced by 30\%.
This result is based on the agreement between BHLUMI and SABSPV for all
four standard event selections, see example in Fig.~\ref{fig:sical92-alf2exp}.
The agreement within 0.1\% between BHLUMI and NLLBHA for the unrealistic
BARE1 trigger was also taken into account.
The estimate of the missing photonic ${\cal O}(\alpha^3 L^3)$ contribution
is based on the result from the new calculation embodied in the
LUMLOG event generator (sub-generator in BHLUMI) \cite{bhlumi4:1996}.
As we see, the older value of this contribution was underestimated.
The light pair contribution was also previously underestimated and
as a result of work and discussion within the working group it went up
to 0.03\%.
There is a condition attached to the use of 0.03\% for pairs -- one has
to use at least the LL calculation of the light pair production effect.
If not, then 0.04\% is recommended.
The new, slightly better value of the vacuum polarization error
is the result of recent works \cite{burkhardt-pietrzyk:1995}
and \cite{eidelman-jegerlehner:1995}.
The corresponding programs for vacuum polarization are included
in BHLUMI \cite{bhlumi4:1996}.
The improved calculation with a new smaller error tag for the Z exchange
was done during the workshop and published in Ref.~\cite{th-95-74}.
The corresponding improvement of the matrix element for the Z exchange
is implemented in BHLUMI \cite{bhlumi4:1996}.

The highest priority is now to calculate ${\cal O}(\alpha^2 L)$
contributions and to implement them in the BHLUMI Monte Carlo 
event generator.
When this correction is under complete control, then the remaining
biggest problem will be to determine and reduce the technical
precision of the Monte Carlo calculation.
(At present, technical precision is combined with the missing
${\cal O}(\alpha^2 L)$ contribution.)

There is an ongoing effort to calculate complete ${\cal O}(\alpha^2 L)$
for the SABH process, see for instance 
Refs.~\cite{arbuzov:1995a,arbuzov:1995b} and the contribution of
L. Trentadue in these Proceedings.
In the next section, we shall present the new unpublished
numerical result for some important  ${\cal O}(\alpha^2 L)$ 
contribution, coming from the Cracow--Knoxville group.

Finally, let us mention only briefly the main results of the
Bhabha Working group on the LABH process.
In the working group the first systematic comparison of 
seven different calculations was done. 
The agreement of 0.5\% close to the Z position was reached
and at LEP2 agreement at the level of 2\% was seen.
Most of the programs were of the Monte Carlo type and the comparison
was made for two realistic ES's, calorimetric
and non-calorimetric, each for two values of the collinearity cut.
The resulting precision estimate essentially confirmed the expectation.
The important achievement was that
it was the first systematic comparison of the LABH programs for
a wide range of event selections.
The result of the comparison is encouraging, but more work
is obviously required if the theoretical error is to be reduced.
We refer the reader for more details to the section on the LABH process
in Ref.~\cite{YR-96-01-bhabhaWG}.

\subsection{New results in the ${\cal O}(\alpha^2 L)$}

\begin{figure}[!ht]
\centering
\setlength{\unitlength}{0.1mm}
\begin{picture}(800,800)
\put( 0, 0){\makebox(0,0)[lb]{
\epsfig{file=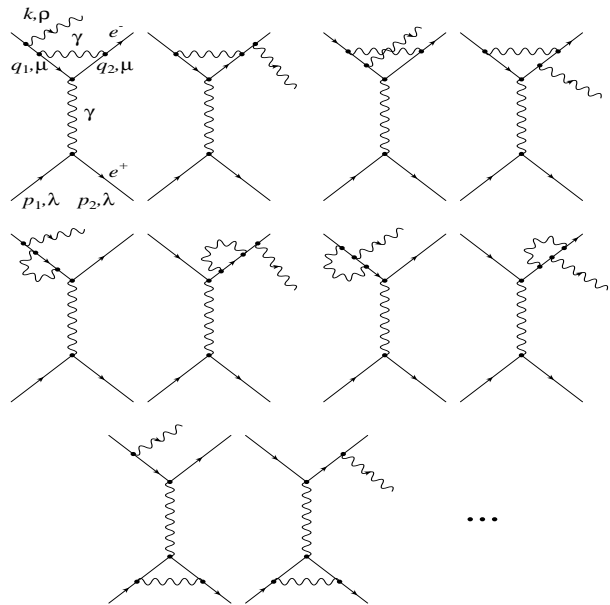,width=80mm,height=80mm}
}}
\end{picture}
\caption{${\cal O}(\alpha^2)$ single bremsstrahlung correction in
$e^+ e^- \rightarrow e^+ e^-$ at low angles. Only the upper line real emission
graphs are shown.}
\label{fig:diagrams}
\end{figure}

\begin{figure*}[!ht]
\centering
\setlength{\unitlength}{0.1mm}
\begin{picture}(1600,800)
\put( 0, 0){\makebox(0,0)[lb]{
\epsfig{file=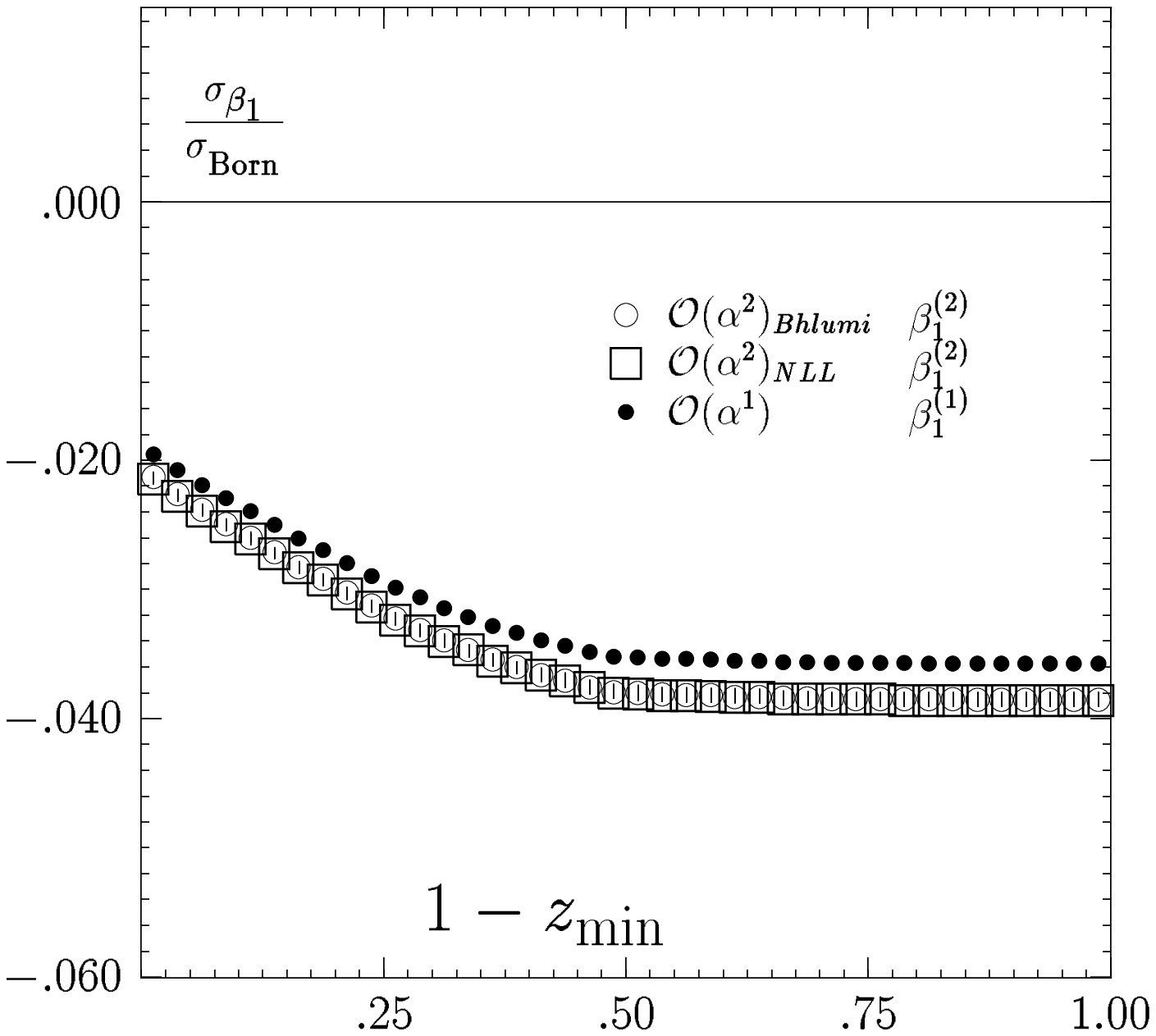,width=80mm,height=80mm}
}}
\put( 800, 0){\makebox(0,0)[lb]{
\epsfig{file=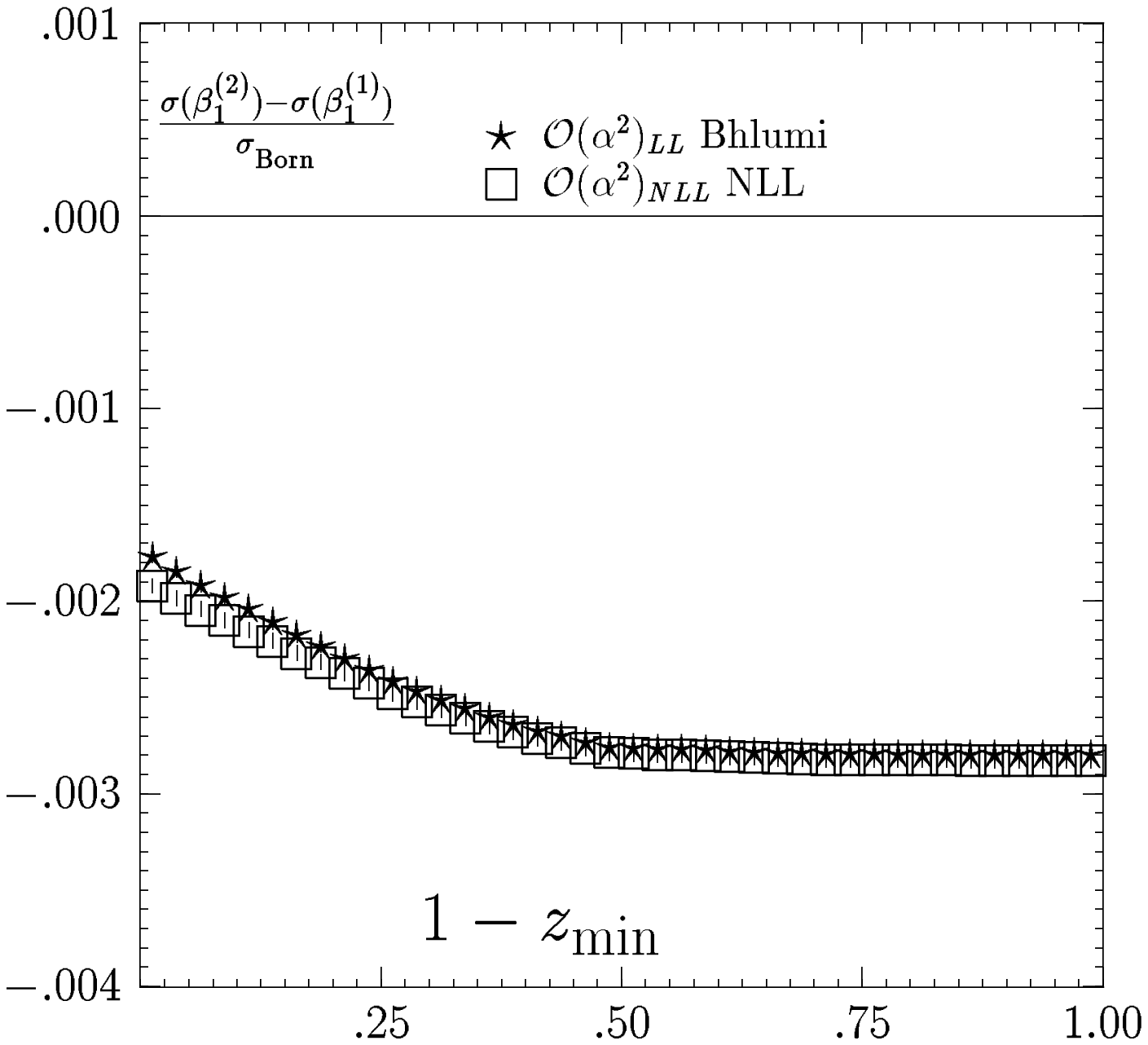,width=80mm,height=81mm}
}}
\end{picture}
\vspace{-8mm}
\caption{
 Monte Carlo result for entire $\bbe_1^{(r)}, r=1,2$
 and for the pure second-order contribution
 $\bbe_1^{(2)} - \bbe_1^{(1)}$
 for the SICAL Wide-Narrow event selection.
 Mass-tems are not included.
 All result divided by Narrow-Narrow Born cross section.
 Energy cut variable $ z_{\min}$ is defined in Fig.~2 
 of Ref.~\protect\cite{th-95-38}.   
}
\label{fig:beta1}
\end{figure*}

The calculation of the second-order QED matrix for the SABH process
includes a calculation of the (a) two-loop second-order form factor,
(b) one-loop correction to single photon emission process and
(c) tree-level double-photon bremsstrahlung.
The calculations of (a) and (c) exist in the literature
and the fully differential distribution for (b) was not available
until recently.
In the following we describe numerical results obtained using
the recently published results of Ref.~\cite{ut-90-0101:1996} on the
one-loop corrections to single-photon emission.
The Feynman diagrams involved in this process are shown in 
Fig.~\ref{fig:diagrams}.
In the BHLUMI event generator this matrix element is implemented
in the \Order{\alpha^2L^2} approximation.
The sub-leading contributions of \Order{\alpha^2L} are incomplete in BHLUMI.
Since the result of Ref.~\cite{ut-90-0101:1996} is in principle {\em exact},
we therefore have a chance to check how big is, in fact,
the missing \Order{\alpha^2L} part.
We expect it generally to be 0.1\% or less.
There is, however, a certain problem from the start.
How can we calculate something meaningful numerically
from the formulas of Refs.~\cite{ut-90-0101:1996} and \cite{th-95-38}
if the matrix element from the Feynman diagrams in Fig.~\ref{fig:diagrams}
taken alone is infrared-divergent.
(We have to combine it with double bremsstrahlung and the second-order part
of the formfactor in order to get a finite cross section.)
The way out that we shall apply for the moment is to subtract both
virtual and real singularities according to Yennie-Frautschi-Suura 
work \cite{yfs:1961}.
The infrared-divergent terms are trivial and uninteresting
-- they are proportional to the perfectly known first-order matrix element.
Only the piece left after the subtraction contains non-trivial \Order{\alpha^2}
corrections; it is finite and small.
It is the so-called $\bbe_1$ function.
The $\bbe_1$ function is a standard object
in the YFS inclusive exponentiation scheme --
it is already used in the BHLUMI matrix element.
It is defined as follows
\begin{equation}
\begin{split}
\bbe^{(2)}_{1}(k_i)
=&  \left\{ D^{(2)}_{[1]}(k)    \exp(-2\alpha\Re B) \right\} 
        \Bigg|_{{\cal O}(\alpha^2)}
 -\tilde{S}_p(k)\; \bbe^{(1)}_0,
\end{split}
\end{equation}
where $D^{(2)}_{[1]}$ is the \Order{\alpha^2} squared matrix element
for single-real-photon emission corresponding to diagrams 
in Fig.~\ref{fig:diagrams}; the definitions of the infrared virtual
formfactor $B$ and of the real soft factor $\tilde{S}$
are exactly the same as in Refs.~\cite{yfs:1961,th-95-38}.
We are really interested in the pure \Order{\alpha^2} part of 
$\bbe^{(2)}_{1}$; we will therefore need to subtract the trivial
first-order version of it defined as
\begin{equation}
\bbe^{(1)}_{1}(k_i)
 =   D^{(1)}_{[1]}(k) -\tilde{S}_p(k)\; \bbe^{(0)}_0.
\end{equation}

The 3-body phase-space integration is done using the BHLUMI Monte Carlo;
the results are shown in Fig.~\ref{fig:beta1}
for the realistic caloric ES called SICAL defined in Ref.~\cite{th-95-38}
as a function of the energy cut $z_{\min}$, also defined there.
As we see in the figure the difference between \Order{\alpha^2}
and \Order{\alpha^1} is about 3\% and therefore compatible
with the generic size of the \Order{\alpha^2L^2}, 
see Table~\ref{tab:canonical-coeff}.
The difference between the BHLUMI LL ansatz of 
Ref.~\cite{th-95-38}\footnote{In the presented results
   we used BHLUMI 4.03 with the corrected matrix.}
and the exact result of Ref.~\cite{ut-90-0101:1996} is up to 0.015\%.
This has to be multiplied by factor 2 because only the emission
from the upper line is taken into account.
We therefore obtain the total missing \Order{\alpha^2L} to be 0.03\%,
i.e. well compatible with expectations.
In the above results result the so-called mass-terms are still 
not included\footnote{Their inclusion should not spoil the agreement 
-- mass-terms are expected to be negligible.}.
They are still under numerical tests.
Note that this result was not known at the time of the LEP2 workshop
and was not taken into account in Ref.~\cite{YR-96-01-bhabhaWG}
and Table~\ref{tab:total-error-lep1}.
The above result shows the most complicated and difficult component
of the \Order{\alpha^2} calculation, but it is still incomplete
because the two-loop formfactor and double-bremsstrahlung contributions
have to be included.
Once all three components are in BHLUMI, we shall be able to make
the theoretical precision of the luminosity measurement better still.
We hope that this will happen soon.
Let us also express our hope that we shall be able to compare our results
with those of Refs.~\cite{arbuzov:1995a,arbuzov:1995b}.

\section{Summary}
We have reported on the recent improvements on the precision of the
QED calculations of the Bhabha process down to the 0.11\% level.
In particular we have shown recent results on the second-order
exact calculation for one real and one virtual photon, which
suggests that missing \Order{\alpha^2} contributions are below 0.1\%.
This opens the way to even further improvements in the theoretical
error on the luminosity measurement at LEP.


\end{document}